\documentclass[5p,twocolumn,times]{elsarticle}
\usepackage{lipsum}
\usepackage{lineno,hyperref}
\modulolinenumbers[5]
\usepackage[pdftex]{color}
\usepackage[font=footnotesize,labelfont=bf]{caption}
\usepackage[font=footnotesize,labelfont=bf]{subcaption}
\journal{Journal of \LaTeX\ Templates}
\usepackage{colortbl}
\usepackage{xcolor}

\usepackage{tikz-cd}
\usepackage{amssymb}

\biboptions{compress}

\usepackage[figuresright]{rotating}

\begin{document}

\begin{frontmatter}


\title{Trade-off between reproduction and mobility prolongs organisms' survival in rock-paper-scissors models}


\address[1]{Institute for Biodiversity and Ecosystem
Dynamics, University of Amsterdam, Science Park 904, 1098 XH
Amsterdam, The Netherlands}
\address[2]{School of Science and Technology, Federal University of Rio Grande do Norte\\
Caixa Postal 1524, 59072-970, Natal, RN, Brazil}
\address[3]{ Edmond and Lily Safra International Institute of Neuroscience, Santos Dumont Institute,
Av Santos Dumont 1560, 59280-000, Macaiba, RN, Brazil}
\address[4]{Department of Computer Engineering and Automation, Federal University of Rio Grande do Norte, Av. Senador Salgado Filho 300, Natal, 59078-970, Brazil}

\author[1,2]{J. Menezes}  
\author[3,4]{E. Rangel} 

\begin{abstract}
We study the spatial rock-paper-scissors model, where resource competitors' cyclic dominance impacts organisms' energy levels. Our model assumes that failed selection interactions can lead to energy loss, reducing the chances of success in the spatial game and hastening decline.
To prevent death by energy insufficiency, organisms of one out of the species strategically perform a trade-off between reproduction and mobility. When prioritising exploring more extensive areas, organisms aim to maximise the chances of acquiring resources to regain high energy levels. 
Through simulation, we examine the effect of survival behaviour on species segregation and spatial patterns. 
Our outcomes show that the trade-off between offspring generation and accelerated movement effectively protects individuals from death due to lack of energy. Moreover, the risk of being eliminated by an enemy in the cyclic game reduces due to the behavioural strategy. Considering a three-state model, we quantify how the trade-off parameter controls the organisms' energy recovery. Computing the median organisms' survival time,
we find that although individuals performing the trade-off strategy may live longer, the organisms of other species are negatively affected by a life expectancy reduction. Our research may elucidate the role of adaptive survival strategies in species persistence and provide valuable insights for ecologists.
\end{abstract}


\end{frontmatter}



\section{Introduction}

There is plenty of evidence that organisms' behaviour plays a central role in the stability of ecosystems \cite{ecology}. Adaptive survival strategies have been reported in many biological systems, where individuals change behaviour to increase their fitness when facing local environmental changes \cite{climatechange,adap2}. For example, the ability to identify hostile regions or lack of natural resources allows animals to flee from enemies and find areas propitious to species proliferation \cite{foraging,BUCHHOLZ2007401}. Additionally, 
other animals perform self-adaptive strategies, adjusting
migratory behaviour in response to internal signals, like individual energy shortage, without needing environmental cues\cite{adap-self,bac-co,bac-co2}. It has also been shown that behavioural strategies 
represent an evolutionary capability which gives species an advantage in the competition in cyclic spatial games and, consequently, affects coexistence probability \cite{adaptive1,adaptive2,Dispersal,BENHAMOU1989375,Causes,MovementProfitable,howdirectional,Agg,Adap,AdapII,adaptive-epl,adaptive-jc,eloi}. 
The knowledge of natural behavioural strategies has also helped the development of recent generations of robots, whose movement emulates the animals' locally adaptive methods \cite{animats}. Many inspired algorithms based on animals' self-adaptive foraging behaviour have been proposed for performance optimisation of computer systems in response to changing conditions \cite{fora-adap-book}.

In this letter, we investigate the spatial rock-paper-scissors model, where organisms may face energy depletion due to failed attempts to conquer natural resources by eliminating other individuals in the spatial game \cite{Coli,bacteria,Allelopathy}. This issue has been addressed in May-Leonard models of two competing species, revealing that the system is stabilised the significant number of deaths due to lack of energy \cite{starvation}. However, researchers have not explored the effects of organisms' adaptive strategies to recover energy to prolong survival. Here, 
we introduce a trade-off between reproduction and mobility performed by individuals of one out of the species whenever needing energy rehabilitation \cite{tradeoff1,tradeoff2,tradeoff3}. This means weak individuals may redirect energy from reproduction to mobility, aiming to maximise the explored area, thus improving the chances of success in the spatial game \cite{Reichenbach-N-448-1046,pa2,combination,tanimoto,sol1,pa1,sol3,sol2}.

Our model considers a three-state energy configuration - high, intermediate, and low. Accordingly, an organism is 
has a $100\%$ chance of winning the competition
in the cyclic spatial game only if at the high-energy level. 
This means that intermediate and low-energy organisms 
may face reversal selection interaction from an individual whose species is inferior in cyclic dominance \cite{reversal}.
During our research, we address the questions: i) how does the
trade-off between reproduction and mobility
impacts the spatial organisms' organisation?; ii) is the self-adaptive trade-off effective in helping individuals recover higher energy levels, thus avoiding death by energy loss?; iii) does the behavioural strategy interfere with individuals' vulnerability to being killed in the spatial rock-paper-scissors game?; iv) 
how the behavioural tactic affects the median organisms' survival time?
\begin{figure}
\centering
\includegraphics[width=45mm]{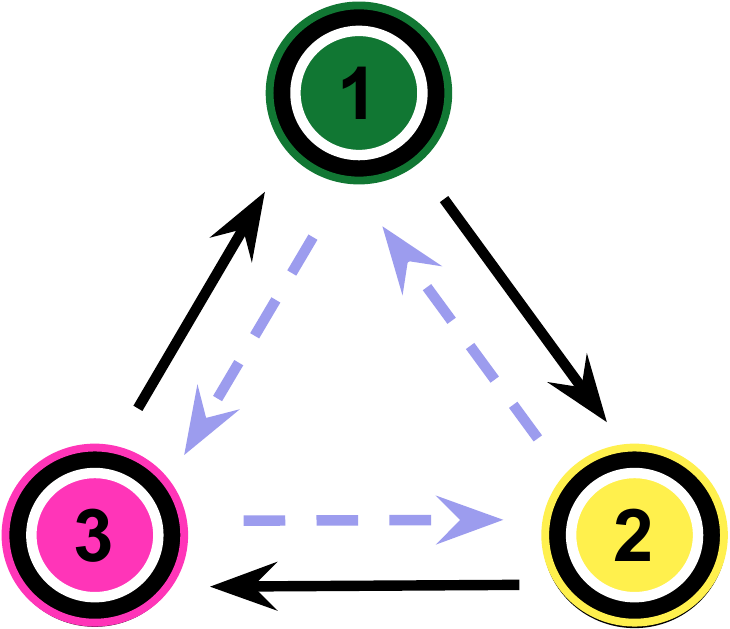}
\caption{Illustration of selection rules in our spatial rock-paper-scissors game model. Solid black arrows indicate the cyclic dominance of individuals of species $i$ over organisms of species $i+1$; the dashed purple arrows show that reversal selection interactions may occur when individuals of species $i$ are in low or intermediate energy states.}
	\label{fig1}
\end{figure}
\begin{figure*}
	\centering
    \begin{subfigure}{.19\textwidth}
        \centering
        \includegraphics[width=34mm]{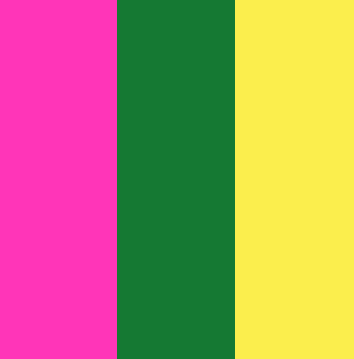}
        \caption{}\label{fig2a}
    \end{subfigure} %
   \begin{subfigure}{.19\textwidth}
        \centering
        \includegraphics[width=34mm]{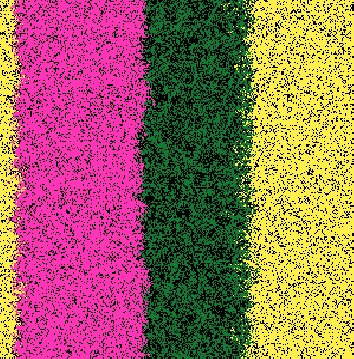}
        \caption{}\label{fig2b}
    \end{subfigure} 
            \begin{subfigure}{.19\textwidth}
        \centering
        \includegraphics[width=34mm]{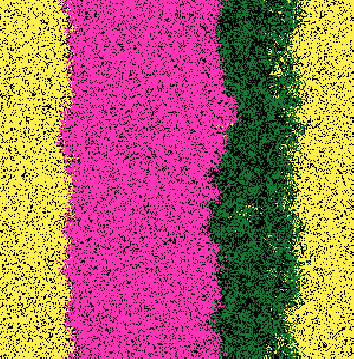}
        \caption{}\label{fig2c}
    \end{subfigure} 
           \begin{subfigure}{.19\textwidth}
        \centering
        \includegraphics[width=34mm]{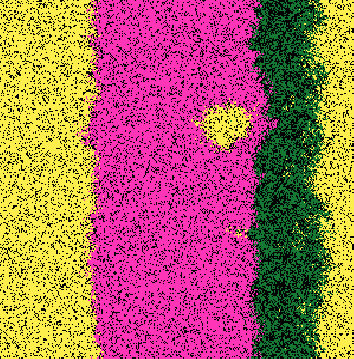}
        \caption{}\label{fig2d}
    \end{subfigure} 
   \begin{subfigure}{.19\textwidth}
        \centering
        \includegraphics[width=34mm]{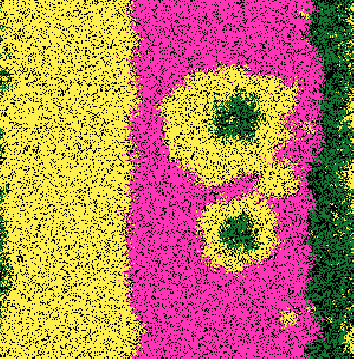}
        \caption{}\label{fig2e}
            \end{subfigure}\\
                \begin{subfigure}{.19\textwidth}
        \centering
        \includegraphics[width=34mm]{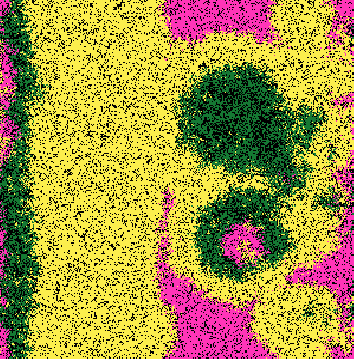}
        \caption{}\label{fig2f}
    \end{subfigure} %
   \begin{subfigure}{.19\textwidth}
        \centering
        \includegraphics[width=34mm]{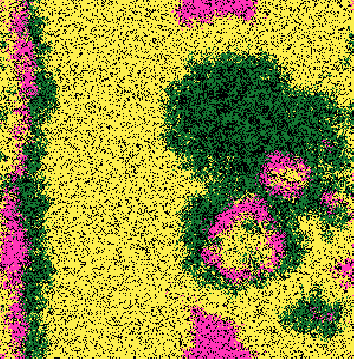}
        \caption{}\label{fig2g}
    \end{subfigure} 
            \begin{subfigure}{.19\textwidth}
        \centering
        \includegraphics[width=34mm]{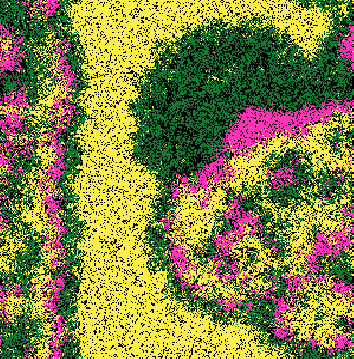}
        \caption{}\label{fig2h}
    \end{subfigure} 
           \begin{subfigure}{.19\textwidth}
        \centering
        \includegraphics[width=34mm]{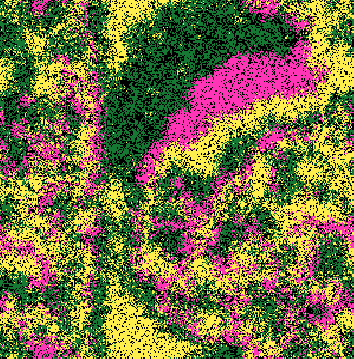}
        \caption{}\label{fig2i}
    \end{subfigure} 
   \begin{subfigure}{.19\textwidth}
        \centering
        \includegraphics[width=34mm]{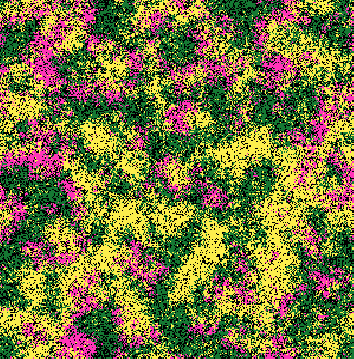}
        \caption{}\label{fig2j}
            \end{subfigure}
 \caption{Snapshots of a simulation of the rock-paper-scissors game starting from the prepared initial conditions in Fig.~\ref{fig2a} 
with the trade-off strategy of individuals of species $1$. 
 in a lattice with $300^2$ grid sites for a timespan of $1000$ generations.
The organisms' spatial organisation at $t=25$, $t=85$, $t=115$, 
$t=160$, $t=200$, $t=260$, $t=305$, $t=405$,
and $t=445$ generations are showed in Figs.~\ref{fig2b}, ~\ref{fig2c}, ~\ref{fig2d}, ~\ref{fig2e}, ~\ref{fig2f}, ~\ref{fig2g}, ~\ref{fig2h}, ~\ref{fig2i},and ~\ref{fig2j}. The colours follow the scheme in Fig~\ref{fig1}; empty spaces appear as black dots. See the spatial pattern dynamics in video https://youtu.be/8eaQEiKyA8M.}
  \label{fig2}
\end{figure*}
\section{The Model}

We investigate a cyclic model of three species that outcompete each other according to the rock-paper-scissors game rules, illustrated in Fig.~\ref{fig1}. The solid black arrows indicate 
that individuals of species $i$ are superior in the cyclic spatial game over organisms of species $i+1$, with $i=1,2,3$, and the cyclic identification $i=i+3\,\alpha$, where $\alpha$ is an integer. 
Our model posits that the organisms of species $i$ can defeat those of species $i+1$ to secure natural resources. But, the organism may experience energy depletion and decreased viability if this is not achieved. When an organism of species $i$ face weakness due to energy loss, the cyclic advantage may be reverted. In this scenario, individuals of species $i$ may be killed by organisms of species $i-1$, as illustrated by the dashed purple arrows in Fig.~\ref{fig1}.

Energy levels are classified into three categories: low (designated as $1$), intermediate (defined as $2$), and high (assigned as $3$). The energy transitioning occurs every time a selection interaction is attempted: in the case of success, the energy level may either increase or maintain the highest level. Conversely, in the event of failure, the energy level may decrease, or individuals with low energy may die.
The organisms' energy level defines the success in the selection interactions:
\begin{itemize}
\item
A high-energy organism of species $i$ always manages to eliminate an individual of species $i+1$, regardless of the defeated organism's energy state.
\item
An intermediate-energy organism of species $i$ experiences reduced strength, compromising its cyclic advantage in competition against high-energy individuals. Therefore, reversal selection is possible, resulting in its elimination by a high-energy organism of species $i+1$.
\item
For low-energy organisms of species $i$, the likelihood of winning against high-energy individuals of species $i+1$ is further decreased. The probability of high-energy individuals of species $i+1$ reversing the selection and eliminating organisms of species $i$ is twice as likely for intermediate-energy levels.
\end{itemize}

In our model, intermediate and low-energy organisms of species $1$ respond to energy depletion by performing a trade-off between reproduction and mobility. The goal is to redirect energy expenditure from producing offspring to increasing the dispersal rate, thus expanding the search region to enhance the likelihood of discovering vulnerable organisms, thus facilitating the recovery of energy levels. This strategy is not motivated by external stimuli but by the need to strengthen personal fitness, to minimise the chances of dying by 
energy depletion. Therefore, we introduce the trade-off
factor, $\beta$, a real parameter ranging from $0$ to $1$, which indicates the proportion of energy redirected from 
reproduction to mobility in response to energy shortage.

\subsection{Stochastic simulations}

Our stochastic simulations run in
square lattices with periodic boundary conditions. We follow the May-Leonard numerical implementation, common to studies of spatial games \cite{leonard,jun,Moura,uneven,appar,hamming,expanding}, 
where the total number of individuals is not conserved. Accordingly, each grid point contains at most one individual; thus, the maximum number of organisms is $\mathcal{N}$, the total number of grid points.

The organisms are initially randomly positioned on the lattice, with each organism occupying a random grid site. The initial conditions are established to have
an equal number of individuals of each species, defined as approximately one-third of the total organisms, or $\mathcal{N}/3$, with $i=1,2,3$. The remaining grid sites are left empty in the initial conditions.

The interactions are stochastically executed using the Moore neighbourhood, where each organism interacts with one of its eight neighbouring individuals. The spatial interactions include:
\begin{itemize}
\item
Selection: an individual of species $i$ eliminates a neighbouring individual of species $i+1$ (direct selection) or $i-1$ (reversal selection), leaving the eliminated individual's grid site empty.
The probability of implementation of a selection interaction depends on the organisms' energy level:
\par
\begin{tikzcd}
\centering
i_l\,\,\,\, (i+1)_{k} \rar{s}\,\, &\,\, i_l \,\,\,\,\, \otimes, \,\,\, l \geq k\,\,\,\,\,\,\,\\
i_l\,\,\, (i+1)_{k} \rar{\eta\,\gamma s/2}\,\, &\,\, i_l \,\,\, \otimes, \,\, k-l=\eta\\
i_l\,\,\, (i-1)_{k} \rar{\eta\,\gamma s/2}\,\, &\,\, i_l \,\,\, \otimes, \,\, l-k=\eta
\end{tikzcd}

where $s$ is the probability of the algorithm sorting a selection interaction to be implemented by an individual of species $i$, and 
$\eta$ is the difference between the energy states of the active and passive individuals. 

The reversal selection is an exception which happens because of the usual winner competitor's weakness. Because of this, the chances of being executed is $\gamma$, a 
real parameter, with $0\leq \gamma \leq 1$, which represents the probability of the 
organism's energy level yields a reversal selection action.
\item
Reproduction: an unoccupied grid site is populated by a new high-energy organism of species $i$. Implementing a reproduction interaction is determined by the probability $r$, for high-energy organisms of every species. However, due to the strategic trade-off, the likelihood of intermediate and low-energy organisms of species $1$ reproducing is reduced to $(1-\beta)\,r$, where $\beta$ is the trade-off factor:
\par
\begin{tikzcd}
\centering
1_l\,\,\,\, \otimes \rar{r}\,\, &\,\, 1_l \,\,\,\,\, 1_l, \,\,\,\,\,\,\,\,\,l=3\\
1_l\,\,\, \otimes \rar{(1-\beta)\,r}\,\, &\,\, 1_l \,\,\, 1_l, \,\,\,\,\, l < 3\\
i_l\,\,\,\, \otimes \rar{r}\,\, &\,\, i_l \,\,\,\,\, i_l, \,\,\,\,\,\,\,\,\, i=2,3
\end{tikzcd}

where $\otimes$ means an empty space and $l$ is the organism energy level.
\item
Mobility: an individual changes location by exchanging places with a vacant space or another organism, regardless of species. Due to the trade-off tactic, the probability of implementing a mobility interaction of intermediate and low-energy energy organisms of species $1$ is higher than the others, namely, $\beta\,r$:
\par\
\begin{tikzcd}
\centering
1_l\,\,\,\, \odot \rar{m}\,\, &\,\, \odot \,\,\,\,\, 1_l, \,\,\,\,\,\,\,\,\,l=3\\
1_l\,\,\, \odot \rar{m\,+\beta\,r}\,\, &\,\, \odot \,\,\, 1_l, \,\,\,\,\, l < 3\\
i_l\,\,\,\, \odot \rar{m}\,\, &\,\, \odot\,\,\,\,\, i_l, \,\,\,\,\,\,\,\,\, i=2,3
\end{tikzcd}

where $\odot$ means an empty space or an individual of any species.
\end{itemize}
The interaction process involves the following steps: i) a random active individual of any species is picked from the organisms in the lattice; ii) one interaction is randomly selected based on the predetermined probabilities; iii) one of the eight neighbouring organisms is randomly chosen to experience the interaction (selection, reproduction, or mobility). Throughout this letter, all results were obtained by assuming the following probabilities: $s = r = m = 1/3$ and $\gamma=0.5$; more, we assume that the probability of a failed 
selection interaction causing organisms' energy decrease is $50\%$.
However, we have verified that our main conclusions hold for other sets of parameters.
Each interaction implementation is recorded as a single time step. After a total of $\mathcal{N}$ time steps, one generation - out time unit - has passed. 


\section{Organisms' spatial organisation}

Let us start by observing the microscopic effects of the trade-off between reproduction and mobility on the organisms' spatial interactions. For this purpose, we perform a single simulation starting from the prepared initial conditions in Fig.~\ref{fig2a}, where individuals are initially allocated in single-species torus rings. This means that each species initially fills one-third
of the grid. The realisation ran in a lattice with $300^2$ grid sites for a timespan of $500$ generations and the following model parameters: $s=r=m=1/3$, $\varepsilon=0.5$, and $\beta=0.9$. 
Figures \ref{fig2b} to \ref{fig2j} show the spatial organisms' organisation at $t=25$, $t=85$, $t=115$, 
$t=160$, $t=200$, $t=260$, $t=305$, $t=405$, respectively. We used the colours in Fig.~\ref{fig1} by depicting individuals of species $1$, $2$, and $3$ with green, yellow, and pink dots; additionally, empty spaces were represented by black dots. The spatial pattern dynamics during the entire simulation in shown in video https://youtu.be/8eaQEiKyA8M.

When the simulation commences, organisms of species $i$ start eliminating individuals of species $i-1$, thus producing a rotation of the rings from left to right. Only organisms in the ring front boundary can access individuals of dominated species; thus, the proportion of individuals failing in the selection interaction is higher far from the border. This leads to individuals' weakening and consequent deaths due to lack of energy, which creates empty spaces, as depicted by black dots inside the single-species spatial domains in Figs.~\ref{fig2b}.

The acceleration of intermediate and low-energy individuals of species $1$ results in an enlargement of the width of the pink torus ring. 
This effect arises from the fact that, although these individuals move faster, their movement lacks directional orientation. Consequently, the rate of individuals from species $1$ entering areas occupied by species $3$ increases; this facilitates the conquest of new territory by individuals of species $3$.

As intermediate and low-energy individuals of species $1$ (green) perform the trade-off strategy, redirecting $90\%$ of effort from reproduction to mobility, the concentration of vacant spaces within the green area becomes higher than inside yellow and pink regions, as observed in Figs.~\ref{fig2b}. 
The consequence is that many individuals of species $2$ manage to infiltrate the green areas because the higher density of vacant sites represents an opportunity to reproduce within the regions of dominant species, as shown in Figs.~\ref{fig2c}. Because newborn individuals of species $2$ are in a high-energy state, they may defeat intermediate and low-energy organisms of species $1$, making it possible for some of them to survive until reaching the pink area, as depicted in Figs.~\ref{fig2d}. From this point on, the proliferation of 
species $2$ (yellow) by eliminating individuals of species $3$ (pink), produces a wave which spreads as shown in Figs.~\ref{fig2e} to ~\ref{fig2g}. The unevenness in the spatial rock-paper-scissors game introduced by the trade-off between reproduction and mobility leads to asymmetric waves with the spatial domains of species $1$ being constantly invaded by individuals of species $2$, as observed in Figs.~\ref{fig2h} to ~\ref{fig2j}. 
\begin{figure}[t]
\centering
    \begin{subfigure}{.49\textwidth}
        \centering
        \includegraphics[width=85mm]{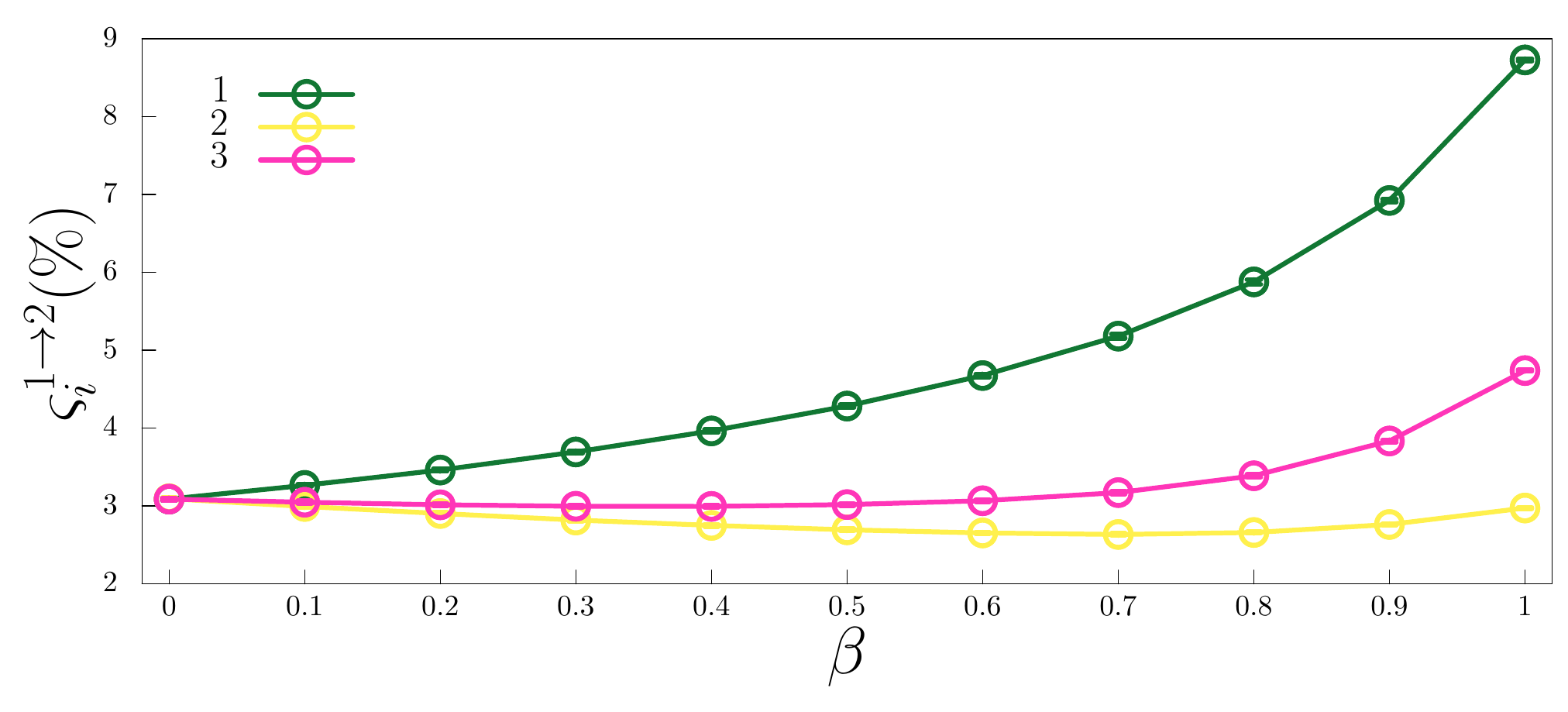}
        \caption{}\label{fig3a}
    \end{subfigure} %
       \begin{subfigure}{.49\textwidth}
        \centering
        \includegraphics[width=85mm]{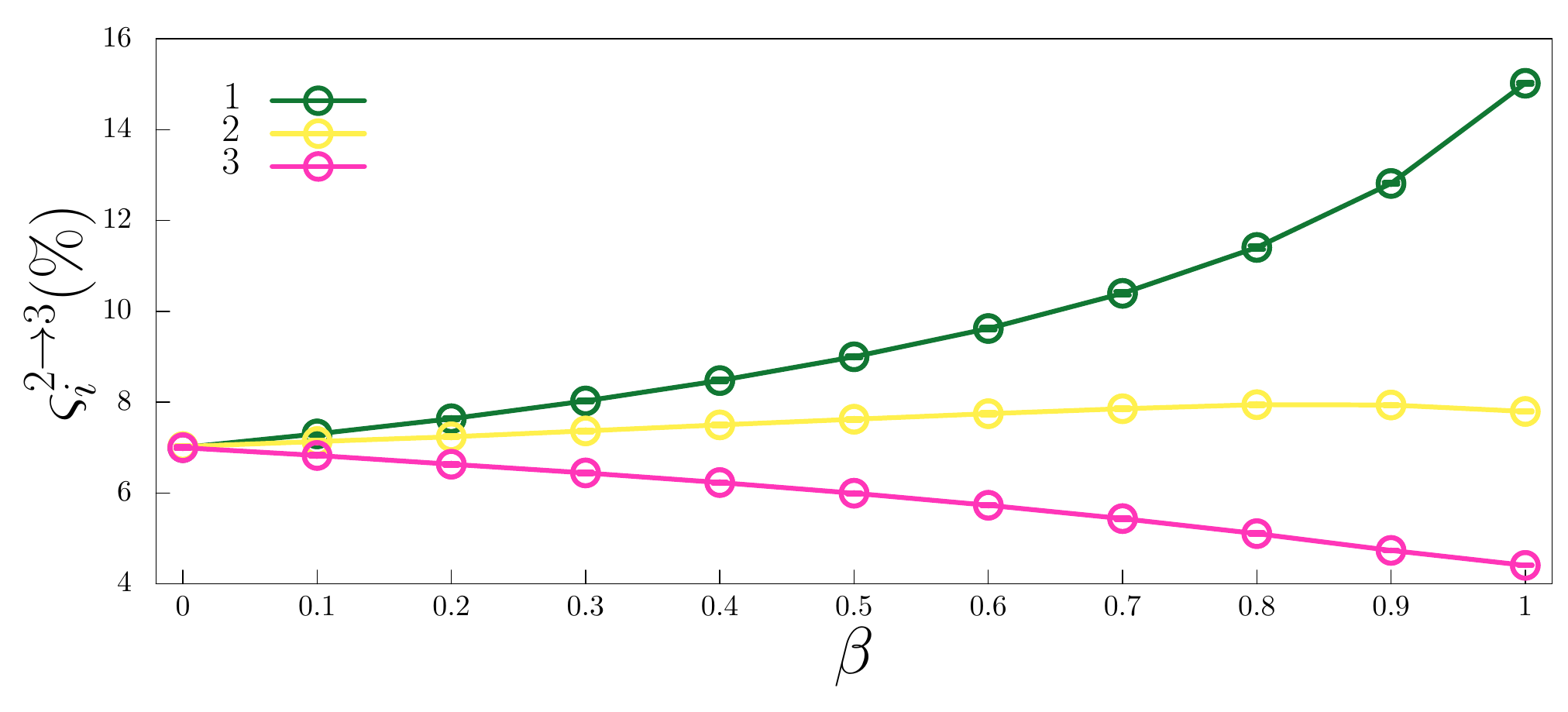}
        \caption{}\label{fig3b}
    \end{subfigure} %
    \caption{Energy recovery rates as a function of the trade-off factor.
Figures \ref{fig3a} and \ref{fig3b} depict the mean value of $\varsigma_i^{1 \to 2}$ and $\varsigma_i^{2 \to 3}$, computed for
set of $100$ simulations; the error bars indicate the standard deviation. The colours follow the scheme in Fig.~\ref{fig1}, where green, yellow and pink stand for species $1$, $2$, and $3$, respectively.}
  \label{fig4}
\end{figure}

\section{Energy recovery rates}

To quantify the effectiveness of the trade-off survival strategy, we compute the chances of the individuals recovering when losing energy. To explore the transitioning of the organisms' energy levels, we introduce the following recovery rates:
\begin{itemize}
\item
$\varsigma_i^{1 \to 2}$: The likelihood of a low-energy individual of species $i$ transitioning to an intermediate-energy state per unit time.
\item
$\varsigma_i^{2 \to 3}$: The probability of an intermediate-energy organism of species $i$ individual change to a high-energy state per unit time.
\end{itemize}

The implementation of these quantities follows the steps: i) counting the number of low and intermediate-energy individuals of species $i$ at the beginning of each generation; ii)
calculating the number of organisms of species $i$ whose energy grows during the generation, distinguishing between intermediate and low-energy ones; iii) $\varsigma_i^{1 \to 2}$ is the ratio between the number of low-energy individuals of species $i$ transitioning to an intermediate-energy state and the number at the beginning of each generation; iv) $\varsigma_i^{2 \to 3}$ is the ratio between the number of intermediate-energy individuals of species $i$ going to a high-energy state and the number at the beginning of each generation.

We performed sets of $100$ simulations starting from different initial conditions for $0 \leq \beta \leq 1.0$, in intervals of $\Delta \beta=0.1$. The simulations ran in lattices with $500^2$ grid sites for a timespan of $5000$ generations. To guarantee the quality of the results, we remove the data from the initial simulation stage, thus calculating the average recovery rates in the second half of each realisation. 
Figures \ref{fig3a} and \ref{fig3b} display the impact of $\beta$ on $\varsigma_i^{1 \to 2}$ and $\varsigma_i^{2 \to 3}$, respectively.
Green, yellow, and pink lines show the average energy recovery rates for individuals of species $1$, $2$, and $3$, respectively, with error bars indicating the standard deviation.

The results show the effectiveness of the trade-off between reproduction and mobility in improving the chances of energy recovery. Comparing the outcomes in Figs.~\ref{fig3a} and ~\ref{fig3b}, one sees that the likelihood of an organism transitioning from low to intermediate energy levels is lower than the chances of transitioning from intermediate to high energy levels, regardless of the species. This happens because low-energy individuals are more susceptible to elimination through reversal selection interactions.

Furthermore, the more significant the proportion of effort redirected to increase the dispersion rate, the higher the probability of energy comeback. For $\beta=1.0$, $\varsigma_1^{1 \to2}$ is at its highest, with low-energy individuals having nearly three times the chances of recovery, compared to the case where the adaptive strategy is absent.

Regarding species $2$ and $3$, whose organisms do not employ the behavioural strategy, our results demonstrate both advantages and disadvantages for energy recovery that depend on the trade-off factor of species $1$. For species $2$, intermediate-energy individuals are helped with increased chances of recovering, reaching a peak at $\beta=0.8$; however, the prospects of transitioning of low-energy individuals worsen for any $\beta$. On the other hand, for species $3$, both intermediate and low-energy are negatively affected if $\beta < 0.7$, but $\varsigma_3^{1 \to2}$ grows for $\beta \geq 0.7$, thereby favouring low-energy organisms.

\begin{figure}[t]
\centering
    \begin{subfigure}{.49\textwidth}
        \centering
        \includegraphics[width=85mm]{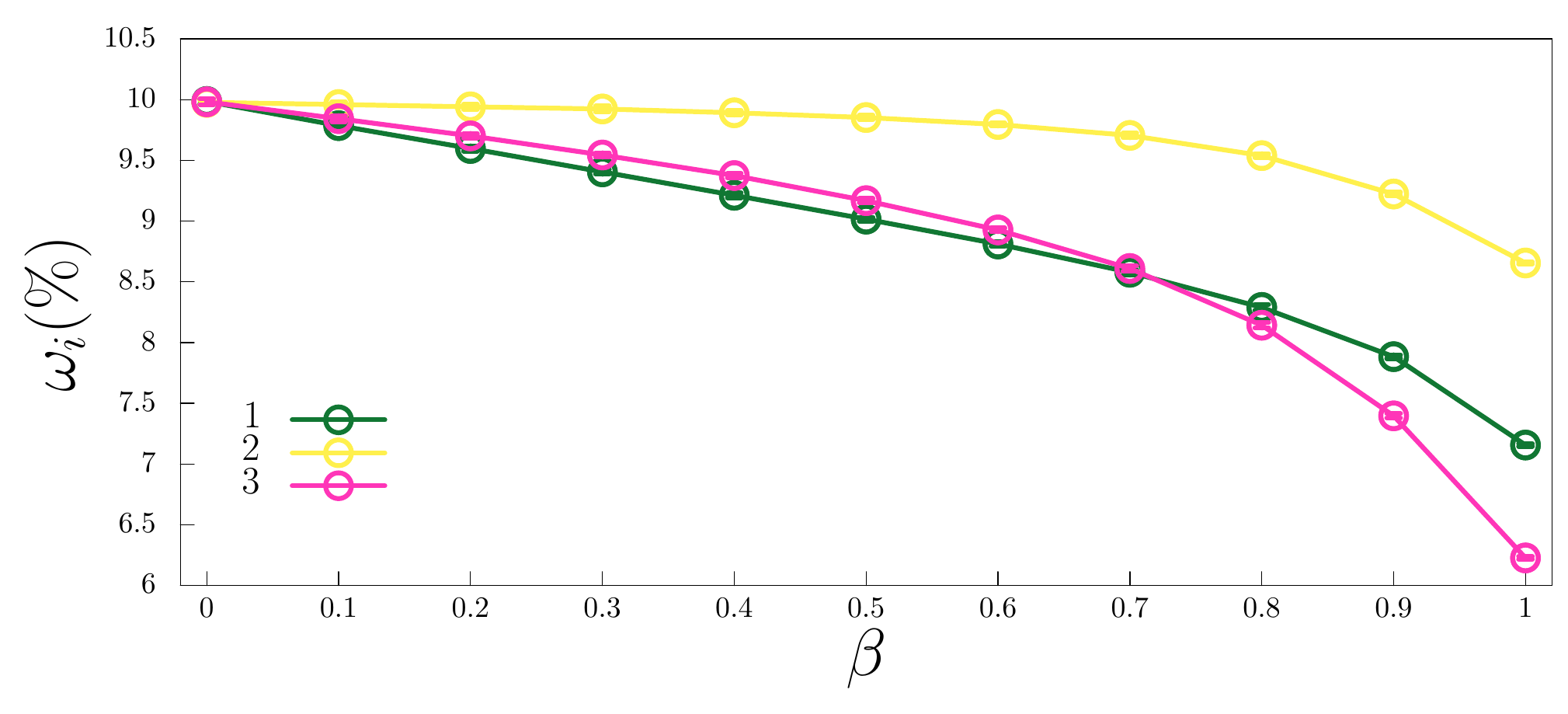}
        \caption{}\label{fig4a}
    \end{subfigure} %
       \begin{subfigure}{.49\textwidth}
        \centering
        \includegraphics[width=85mm]{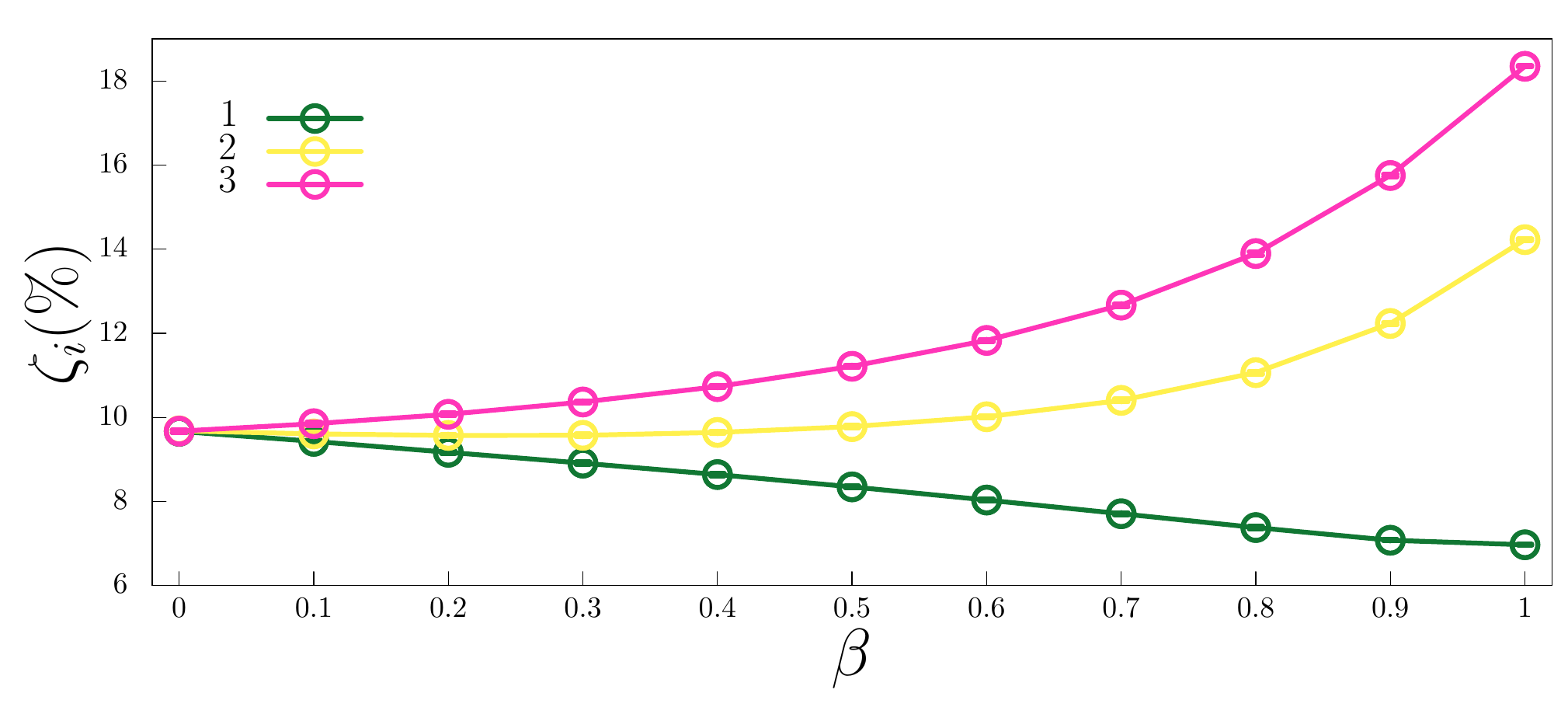}
        \caption{}\label{fig4b}
    \end{subfigure} 
    \caption{Depletion and selection risks in terms of the trade-off factor. $\omega_i$ and $\zeta_i$ were calculated through collections of $100$ simulations, are shown in Figs. \ref{fig4a} and \ref{fig4b}; the error bars indicate the standard deviation. The colours follow the scheme in Fig.~\ref{fig1}.
}
  \label{fig4}
\end{figure}

\section{Depletion and selection risks}

We now investigate how the trade-off strategy 
influences the organisms' survival probability of species $i$. 
For this purpose, we first calculate the death rate, differentiating deaths by energy insufficiency from selection in the spatial game.

First, we introduce the depletion risk $\omega_i$: the probability of an organism of species $i$ perishing due to weakness per unit time. This is implemented as follows:
i) counting the number of individuals of species $i$ at the beginning of each generation (irrespective of the energy level); ii) computing how many individuals of species $i$ die because of energy lack during the generation; iii) computing the depletion risk as the rate of the number of dead individuals and the total number at the beginning of each generation.

Second, we compute the selection risk $\zeta_i$: the probability of an organism of species $i$ being eliminated in the spatial rock-paper-scissors game per unit time. The implementation follows the steps: i) counting the number of individuals of species $i$ at the beginning of each generation (irrespective of the energy level); ii) computing how many individuals of species $i$ are killed by adversaries in the spatial game during the generation; iii) computing the selection risk as the rate of the number of killed individuals and the total number at the beginning of each generation.

We use data from collections of $100$ simulations starting from different initial conditions for $0 \leq \beta \leq 1.0$, in intervals of $\Delta \beta=0.1$. The simulations ran in lattices with $500^2$ grid sites for a timespan of $5000$ generations.
We removed the fluctuations in data that occur during the formation of spatial patterns in the initial stage of simulations by utilising the average depletion and selection risks found from the second half of each simulation run.
Figures \ref{fig4a} and \ref{fig4b} depict the impact of $\beta$ on $\omega_i$ and $\zeta_i$, with green, yellow, and pink lines representing species $1$, $2$, and $3$, respectively.; the error bars indicate the standard deviation. 

The outcomes unveil that the survival tactic employed by individuals of species $1$ effectively protects against energy loss-related fatalities and from enemy attacks in the cyclic game, as depicted by the green lines in Figs.~\ref{fig4a} and ~\ref{fig4b}, respectively. As $\beta$ grows, individuals of $2$ and $3$ also profit from the strategic behavioural strategy of individuals of species $1$: the chances of death due to lack of energy decrease. However, they become more vulnerable to dying by being caught by enemies, as shown by the yellow and pink lines in Fig.~\ref{fig4b}.

\section{Median survival time}

Finally, we compute the estimated organisms' median survival time using the survival probability as a function of the trade-off factor
\begin{equation}
\mathcal{S}_i (\beta)\,=\,1\,-\,\omega_i (\beta)\,-\,\zeta_i (\beta).
\end{equation}
The median survival time of individuals of species $i$, $\mathcal{T}_i$, is given by assuming the threshold $\mathcal{S}_i^{\mathcal{T}} \,=\,0.5$.

\begin{figure}[t]
	\centering
	\includegraphics*[width=8.8cm]{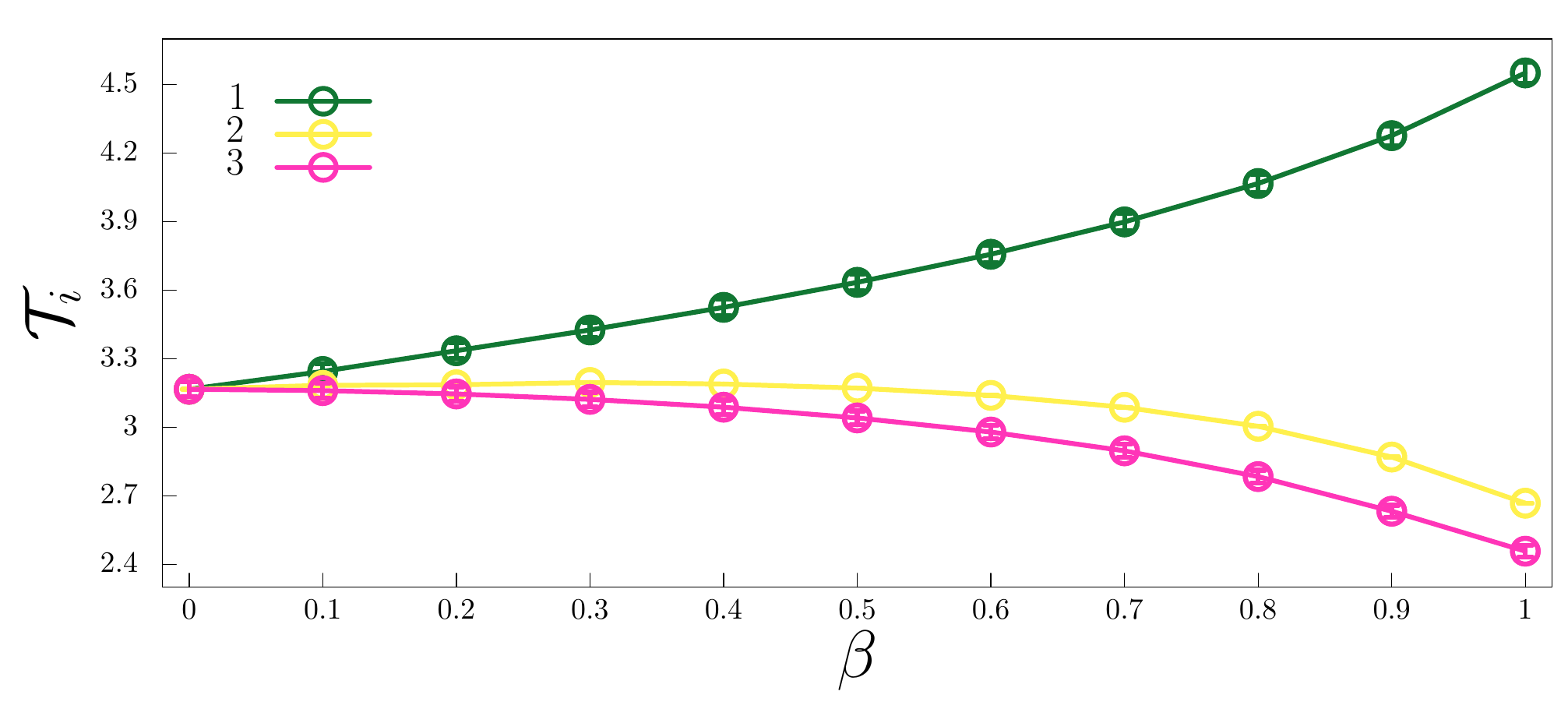}
\caption{
Estimated organisms' median survival time as a function of the trade-off factor. The results were obtained 
running collections of $100$ simulations in lattices with $500^2$ grid sites, running until $5000$ generations; the error bars indicate the standard deviation. The colours follow the scheme in Fig.~\ref{fig1}.}
 \label{fig5}
\end{figure}

Figure \ref{fig5} shows the estimated organisms' median survival time in terms of the trade-off factor, averaged from sets of $100$ simulations in lattices with $500^2$ grid sites, running until $5000$ generations; the error bars indicate the standard deviation. Green, yellow, and pink lines depict the results for species $1$, $2$, and $3$, respectively. 
Accordingly, organisms of species $1$ live longer as 
more energy is redirected from reproduction to mobility: the maximum relative growth in $\mathcal{T}_1$ is $44\%$, reached for $\beta=1.0$. In contrast, the estimated median survival time of individuals of species $3$ significantly reduces, with organisms living $29\%$ less, on average, for $\beta=1.0$. In the case of species $2$, $\mathcal{T}_2$ slightly grows for $\beta \leq 0.5$, with the relative increase reaching $0.72\%$ for $\beta = 0.5$; however, the median time survival of individuals of species $2$ drops for $\beta > 0.5$, with the maximum relative decrease, $19\%$, occurring when intermediate and low-energy organisms of species $1$ redirect to mobility the total energy usually spent in reproduction activity.

\section{Comments and Conclusions}

Studying the stochastic version of the spatial rock-paper-scissors game, we explored the benefits of the trade-off between mobility and reproduction on organisms' energy levels and survival. For this, we consider a three-state energy approach, where individuals are weakened when they fail to take the resources from an adversary in the cyclic game. This provokes the organisms' weakening, leading to a vulnerability in playing the spatial game, thus accelerating the demise by energy depletion. 

Our simulations showed that if intermediary and low-energy organisms prioritise mobility over reproduction, the survival probability increases because: 
\begin{itemize}
\item
the success rate in the cyclic spatial competition rises, thus being more likely to regain energy;
\item
the risk of being eliminated by an adversary in the cyclic game drops.
\end{itemize}

Our discoveries show that the trade-off strategy profoundly impacts the spatial organisms' organisation, with the deformation of the single-species domains common to unbalanced spatial rock-paper-scissors models.  Although other species also benefit from maintaining high-energy levels, they do not profit from protection against enemies. The outcomes show that if individuals of species $i$ redirect energy from reproduction to mobility, their chances of organisms of other species being caught by an enemy in the cyclic game increase, with individuals of species $i-1$ being the most vulnerable to dying in the cyclic game. Because of this, only the estimated survival time of organisms of species $i$ is prolonged.

The evolutionary trade-off strategy can significantly impact the average organisms' survival time, with benefits for one species but adverse effects for others. 
Therefore, this adaptive strategy, which aims to minimize the death risk, directly interferes with biodiversity stability. Further study can explore the role of organisms' conditioning in coexistence probability, focusing on the potential trade-offs between reproduction and mobility, which could promote or jeopardise biodiversity. 

\section*{Acknowledgments}
We thank CNPq, ECT, Fapern, and IBED for financial and technical support.

\bibliographystyle{elsarticle-num}
\bibliography{ref}

\end{document}